# Quasars with Super Metal Rich Emission Line Regions

by

Neelam Dhanda[1,2], Jack A. Baldwin[1,2],

Misty C. Bentz[3] and Patrick S. Osmer[3]

**Abstract**

We study the degree of chemical enrichment in the Broad Emission Line Regions (BELRs) of two QSOs with unusually strong nitrogen emission lines. The N V λ1240/ C IV λ1549 intensity ratio is often used as a metallicity indicator for QSOs. The validity of this approach can be tested by studying objects in which the N IV] and N III] lines, in addition to N V, are unusually strong and easily measurable. If all of these ionization states of nitrogen point to the same metallicity, it implies that the large N V strengths observed in most QSOs are not due to some peculiarity of the N V λ1240 line. This test had previously been applied to Q0353-383, a QSO long known to have extremely strong N III] and N IV] lines, with the result supporting high metallicity in that object. Here we make the same check in two other QSOs with very strong nitrogen lines, as a step towards using such QSOs to better probe the early chemical enrichment histories of their host galaxies. J1254+0241 has a metallicity of about 10× solar, with good agreement between the abundance results from different line ratios. J1546+5253 has a more moderate metallicity, about 5× solar, but the abundances determined from different line ratios show a much wider scatter than they do for J1254+0241 or Q0353-383. This QSO also has an unusual low-ionization emission line spectrum similar to some low-ionization BAL QSOs and to the unusual AGN I Zw 1. We attribute the peculiarities in its spectrum to some combination of unusual structure and/or unusual physical conditions in its BELR. Our results further affirm the validity of the N V/C IV ratio as an abundance indicator in QSOs.

*Subject headings:* quasars: emission lines, galaxies: active

## 1. Introduction

The luminous quasars seen at high redshifts occur in the cores of recently-formed massive galaxies (Richstone et al. 1998; Gebhardt et al. 2000; Ferrarese & Merritt 2000). The broad emission line regions in these objects contain several thousand solar masses of gas (Baldwin et al. 2003a) that presumably has fallen in from the inner regions of the surrounding galaxy. The chemical abundances in these Broad Emission Line Regions (BELRs) should therefore reflect the degree of chemical enrichment in the inner bulges of the host galaxies, and thus can tell us about the early history of chemical enrichment of massive galaxies.

[1] Physics and Astronomy Department, Michigan State University, 3270 Biomedical Physical Sciences Building, East Lansing, MI 48824; dhanda, baldwin@pa.msu.edu

[2] Visiting Astronomer, Kitt Peak National Observatory, National Optical Astronomy Observatory, which is operated by the Association of Universities for Research in Astronomy, Inc. (AURA) under cooperative agreement with the National Science Foundation.

[3] Department of Astronomy, The Ohio State University, 140 West 18th Avenue, Columbus, OH 43210; bentz, osmer@astronomy.ohio-state.edu

Over the past decade a technique has been developed to measure the overall metallicity $Z$ of these BELRs, where $\log (Z/Z_\odot) \approx \log(O/H) - \log(O/H)_\odot$ with O and H indicating the abundances of these elements by number. The approach (Hamann & Ferland 1993, 1999; Hamann et al. 2002) is to directly measure the N/O and N/C abundance ratios, and then use the strong correlation between these two ratios and $Z$ that is observed in our own Galaxy (e.g. Pettini et al. 2002) to determine $Z$. The physical origin of the observed correlation is the buildup of N as a secondary product in CNO cycling in higher mass stars. This approach for measuring BELR chemical abundances is needed because it is impossible to directly measure the O/H or C/H abundance ratios, due to the fact that the H emission lines record the heating rate in the BELR gas while the lines of the elements heavier than He (including C, N and O) record the cooling rate, and the heating and cooling rates always adjust to be in balance. The Hamann & Ferland technique instead compares the strengths of emission lines that are competing to carry the cooling load, including lines of C, N and O. When the abundance of N is higher relative to C and O, lines of N are able to carry a greater fraction of the cooling, at the expense of the collisionally excited lines of the other elements. This makes the N/O and N/C line ratios sensitive to the relative abundances of these elements.

Often the only N line that is strong enough to be measured in QSO spectra is N V $\lambda$1240, which is heavily blended with Ly$\alpha$. As was first shown by Hamann & Ferland (1993), the strength of this line relative to those of C IV $\lambda$1549, He II $\lambda$1640 and other lines consistently indicates quite high metallicities in the BELRs of many QSOs, in the range of several times the solar metallicity. The metallicity is often described as correlating with QSO luminosity (e.g. Hamann & Ferland 1993; Shemmer & Netzer 2002; Dietrich et al. 2003; Nagao, Marconi & Maiolino 2006), although the true correlation is likely to be with some combination of the mass and accretion rate of the central black hole (Warner et al. 2003, 2004; Shemmer et al. 2004). High metallicities are in fact expected; simple chemical evolution models show that in the environment of the central region of a massive galaxy $Z$ can build up to 10–20 times the solar value in the space of only a Gyr or so (Hamann & Ferland 1999; Friaca & Terlevich 1998; Romano et al. 2002).

However, measuring the N/O and N/C abundance ratios depends on fitting the observed line intensity ratios to the predictions of models of photoionized BELRs that are at least approximately correct. It is important to try to also use lines from other ionization states of N, in addition to $N^{+4}$, in order to check for errors due to overlooked ionization or excitation effects as well as for errors in measuring the heavily blended N V line. Intensity ratios involving $N^{+2}$ and $N^{+3}$ intercombination lines are also expected to be useful abundance indicators (Shields 1976; Hamann et al. 2002). The metallicities derived from these lines have been compared to those found from N V both for samples of individual QSO spectra (Dietrich & Wilhelm-Erkens 2000; Shemmer & Netzer 2002; Dietrich et al. 2003; Shemmer et al. 2004) and for sets of composite spectra each representing many QSOs binned together by redshift, luminosity, or central black hole parameters (Warner et al. 2003, 2004; Nagao et al. 2006). These studies generally show good agreement between the metallicities derived from the intensity ratios $I$(N V $\lambda$1240)/$I$(C IV $\lambda$1549), $I$(N V $\lambda$1240)/$I$(O VI $\lambda$1034), $I$(N V)/$I$(O VI $\lambda$1034 + C IV $\lambda$1549), $I$(N IV] $\lambda$1486)/$I$(O III] $\lambda$1664) and $I$(N III] $\lambda$1751)/$I$(O III] $\lambda$1664). These all indicate metallicities agreeing to better than a factor of 2 when averaged over fairly large samples of objects. Other intensity ratios ($I$(N V $\lambda$1240)/$I$(He II $\lambda$1640), $I$(N IV] $\lambda$1486)/$I$(C IV $\lambda$1549) and $I$(N III] $\lambda$1751)/$I$(C III] $\lambda$1909)) suggested by Hamann et al. (2002) show a clear correlation with the above abundance indicators, but with larger scatter and/or systematic offsets which can at least in part be explained by additional dependences on the ionizing continuum shape or the average gas temperature or density.

These tests of the validity of the metallicity indicators depend on measurements of lines that are extremely weak and blended in many of the spectra. In addition, the composite spectra used in some of those studies average together individual spectra which do not all include every emission line of interest. Given these difficulties, there is great value in repeating such tests for individual bright QSOs in which the N III] and N IV] lines are unusually strong and therefore can be measured accurately. That is the approach we take here. Assuming that the abundance indicators work, these same objects also are representative of the very highest metallicity galaxy cores in the high redshift universe, and therefore can be used to test chemical enrichment models.

Baldwin et al. (2003b) carried out this check for Q0353-383, the one QSO known at that time to have very strong and easily measurable lines of $N^{+2}$ and $N^{+3}$ in addition to the usual $N^{+4}$ line. This was considered to be a "torture test" for this method of measuring $Z$. In this object, the lines of all the different observed N ionization states did turn out to imply similar values of $Z$, corresponding to $Z \sim 15 Z_\odot$. While this appears to validate the basic technique for measuring metallicity, the very high value of $Z$ raises many interesting questions about the nature of Q0353-383. One possibility is that this object has been caught just as the metallicity in the central part of the host galaxy has peaked, at the point where the interstellar gas supply is nearly exhausted and hence the fuel source for the central QSO is ready to shut off.

Q0353-383 was originally discovered to have strong N III] and N IV] lines by Osmer & Smith (1980) as part of a general objective prism search for new QSOs, and its unusual chemical properties were quickly studied by Osmer (1980). There then followed a period of two decades in which this was the *only* QSO known to have such strong N III] and N IV] lines, out of many thousands of QSOs for which spectroscopic data were available. Others have now turned up in massive surveys of the sky such as the Sloan Digital Sky Survey (SDSS). Recently, Bentz, Hall & Osmer (2004) searched the spectra in the SDSS First Data Release (DR1) for additional QSOs with extremely strong N III] and N IV] lines. Of these, four appeared to have nitrogen emission as strong as or stronger than that seen in Q0353-383. In this paper we report on follow-up spectroscopy of these four best candidates. Our goals are to carry out a further check on the validity of the N-line technique for measuring $Z$, and to begin to map out the general properties of these "super metal rich" QSOs in order to understand how they fit into the general picture of the early evolution of massive galaxies that is shown by the larger surveys of QSO metallicity described above. Also, as will become clear below, some of these N-loud objects appear to be similar to Low-ionization Broad Absorption Line (LoBAL) QSOs, and may carry important information about the general structure of the BELR.

## 2. Observations and preliminary data reduction

The observations were taken with the 4m Mayall Telescope at Kitt Peak National Observatory. Table 1 is the observing log. Hereafter we will designate the four QSOs listed in Table 1 with their shortened names: J0909+5803, J1254+0241, J1546+5253 and J1550+0236. We used the RC spectrograph for four nights (28–31 May 2006 UT). We used grating KPC10a for the first three nights to obtain 6.4 Å FWHM resolution spectra of all four objects (but most of the first night was lost because of high winds). These spectra are useful over the observed wavelength range $\lambda\lambda 3200$–6400 Å, and were designed to include the Ly$\alpha$+N V blend as well as all other emission lines to wavelengths well beyond the $\lambda 1909$ blend. The conditions were photometric on the second and third nights. Most of the QSO observations were made through a 1.5 arcsec wide slit taking many exposures typically 1800s in length. Shorter (usually 900s) observations of each QSO were also made through a 6 arcsec slit, and three standard stars were observed each night through the same wide slit, to provide accurate spectrophotometric calibrations. In all cases the slit was oriented at the parallactic angle.

J1254+0241 and J1546+5253, the two QSOs which these low resolution spectra showed to have the strongest N III] and N IV] lines, then were observed on the final night with grating KPC-007. This setup gave 3.2 Å resolution over the wavelength range $\lambda\lambda 3866$–6763 Å (rest wavelength range $\lambda\lambda 1350$–2250Å in these two objects). The higher resolution was intended to help with the deblending and fitting of the many overlapping emission lines in this spectral region. The slit widths and observing procedure were the same as with the lower resolution grating.

The data were reduced using mostly the standard IRAF[4] packages, the exception being that we used a special routine to detect and interpolate over cosmic rays in the two-dimensional images. For the narrow slit spectra, the one-dimensional spectra were extracted using the 1D optimal extraction option in the IRAF *apextract* package. Each of the narrow slit spectra was put onto a linear wavelength scale using an arc lamp spectrum taken before the start of the night, and then the measured wavelengths of strong night-sky emission lines in the spectra were used to make slight shifts in the wavelength scale to accurately remove the effects of instrumental flexure. For each object, all of the narrow-slit spectra were co-added without any flux calibration or atmospheric extinction corrections having been applied. This weights each spectrum by the total number of counts recorded at each wavelength, which gives the optimal signal-to-noise *(S/N)* ratio in the final co-added spectra.

The wide-slit spectra (including those of the standard stars) were extracted using the normal summation over a fixed aperture, and were wavelength calibrated and corrected for atmospheric extinction using the usual IRAF routines. The flux calibration was worked out from the standard star spectra and applied to the wide-slit observations of the QSOs. We then applied the same flux calibration to the co-added, narrow-slit QSO spectra, using an average airmass for the extinction correction. This produced narrow-slit spectra that were approximately flux calibrated, but for which the atmospheric extinction correction is not quite right and which are affected by light loss due to the narrow slit. We divided the flux-calibrated wide-slit QSO spectra by these roughly calibrated narrow-slit QSO spectra and fit the resulting ratio spectra with quadratic functions in wavelength. This produced smooth wavelength-dependent curves which could be used to correct the narrow-slit spectra to the same flux scale as the wide-slit spectra. In this way, we obtained fully calibrated narrow-slit spectra of these rather faint QSOs. We believe these flux values to be accurate to about 10 percent.

The final steps were to correct these spectra for reddening inside our own Galaxy (using the Schlegel et al. 1998 values for $A_B$ and a total-to-selective extinction ratio $R = 3.1$) and to convert the wavelengths to the rest frame using the redshift $z$ values given in Table 1. We used the standard IRAF tasks to do this. The redshift of J0909+5803 was found from the Fe II $\lambda$1787 line, while those of J1254+0241 and J1550+0236 were measured from the C IV $\lambda$1549 emission line. In the case of J1546+5253, the C IV profile is chopped up by absorption lines, so we measured an average redshift from the peak wavelengths of several emission lines: Ly$\alpha$, C II $\lambda$1335, He II $\lambda$1640, N IV] $\lambda$1486 and C IV $\lambda$1549.

Figure 1 shows the fully reduced low-resolution spectra for all four objects. In this paper, we further analyze only the spectra of the two objects with best *S/N* ratio, J1254+0241 and J1546+5253, which are also the ones for which we have both low- and high-resolution spectra. The high resolution spectra for these two objects are shown in Figure 2.

We compared our reddening-corrected spectra to the reddening-corrected SDSS spectra that are publicly available from the SDSS web site[5]. We found that in order to match them to the continuum level of our KPNO spectra at rest wavelength 1830Å, the SDSS spectra had to be multiplied by the following factors: J0909+5803, 1.44; J1254+0241, 1.08; J1546+5253, 1.90; J1550+0236, 2.13. After multiplying the full SDSS spectra by those constants, they generally agree with our results to within the noise at all wavelengths, so that the relative intensities of different emission lines would be the same as measured from either data set. The major exception is that for J1550+0236 the SDSS spectrum curves upwards relative to the KPNO spectrum at its blue end ($\lambda_{observed} <$ 4100Å) ends, suggesting an atmospheric dispersion effect. The other case where there are significant differences is that for J1546+5253 all of the emission lines appear to have about 60 percent larger equivalent width in the SDSS spectrum than in our spectrum, so that the emission lines in the two spectra agree to about 20 percent. This suggests that the continuum varied between the two observations. Given these various discrepancies, we went back and carefully rechecked the flux calibrations for the KPNO data used here, and believe that they are correct.

---

[4] IRAF is distributed by the National Optical Astronomy Observatories, which are operated by the Association of Universities for Research in Astronomy, Inc., under cooperative agreement with the National Science Foundation.

[5] www.sdss.org

## 3. Measurements of the emission line strengths

The next step was to fit and subtract templates of Fe II emission. A grid of such templates based on the Vestergaard & Wilkes (2001) study of I Zw 1, but broadened in velocity by different amounts, was kindly provided to us by M. Vestergaard. These particular templates do not include Fe III. For each QSO, we used the template with the broadening which most closely matched the measured widths of strong and relatively unblended emission lines (N IV] or C IV). To the extent possible, the fit was guided by the strength of the "UV bump" in the $\lambda\lambda 2240$–$2650$Å rest wavelength region, where the Fe II emission normally is strongest. The individual line profiles described and shown below have had this Fe II emission removed before any further analysis was done. None of these QSOs have very strong Fe II bumps, so this correction was modest in all cases. The Fe II $\lambda 2450$ strengths relative to the continuum, as defined in Hartig & Baldwin (1986), are 0.07 for Q1254+0241 and 0.08 for Q1546+5253. These values are about half the median values for BAL QSOs and are typical of the values found for normal QSOs.

We then measured the emission-line strengths by finding and isolating individual lines which we could use as template velocity profiles for fitting to weak and/or blended lines. This is the same technique that was used by Baldwin et al. (2003b) to measure the line strengths in Q0353-383. It has the advantage that it uses empirically determined line profiles, which in real life exhibit a wide variety of non-Gaussian shapes. It has the disadvantage that different lines in the same QSO can have quite different profiles, so the template profile often does not provide an exact fit. The initial fits were done by eye, and then the exact line strengths were finalized automatically using a $\chi$-squared minimization technique.

Table 2 lists the results of this profile fitting, with the fluxes in units of the total flux in the C IV $\lambda\lambda 1548.20, 1550.78$ doublet. The table also lists minimum and maximum fluxes for each line which, except as noted below, are based on the best fit with alternate template profiles which did not fit as well as the best-fitting template. In this or any other technique for measuring the strengths of broad, blended QSO emission lines, the errors usually are not due to the statistics of the fit, but rather are completely dominated by the uncertainties in systematic effects such as the true shape of the line profile, whether or not additional weak lines should be included in the blend, and the level and shape of the underlying continuum. The maximum and minimum fluxes listed in Table 2 are our best estimates of these effects. Table 2 also lists total fluxes for several of the strongest features, measured by integrating the flux above the continuum for the Ly$\alpha$+N V+Si II blend (called "Ly$\alpha$ blend"), N IV] ("N IV] total"), the C IV $\lambda\lambda 1548.20, 1550.78$ blend ("C IV total"), the N III] blend ("N III] total"), and the feature including Al III, Si III], C III] and Fe III ("$\lambda 1909$ blend"). The two measuring techniques give results that are in good agreement.

### 3.1 J1254+0241

This QSO has a fairly normal-looking spectrum with the exception that the N lines are unusually strong. We isolated three different template profiles that appeared to have real differences between their shapes. These were N IV] $\lambda 1486$, C IV $\lambda 1551$ and He II $\lambda 1640$. The N IV] profile was essentially unblended, so we applied a modest smoothing (by fitting a high order polynomial to the full spectrum) and then extracted the profile from the low dispersion data. The measured flux quoted for this line in Table 2 is for fitting the smoothed profile back to the original unsmoothed data. The upper limit is from fitting the template profile to the same line in the high-resolution spectrum, and shows that there is a 20 percent discrepancy between the two spectra at this wavelength, which we believe comes from differences in the polynomial fits used in the flux calibrations for the two different sets of data.

The C IV line is a closely spaced doublet blend ($\lambda\lambda 1548.20, 1550.78$). Our approach for starting with the observed blend and iteratively separating it into its two components is described by Baldwin et al. (1996). The result is a single template profile (called "C IV" in the "Template" columns of Table 2), which accurately reproduces the observed blend if combined together as two lines with the correct wavelength separation and intensity ratio. The flux listed for C IV λ1549 in Table 2 is for the blend of the two lines, not for the individual component. However, the individual component profile is used for fitting to other lines.

The He II λ1640 line also occurs in a blend. The Fe II subtraction produced a stretch of what appears to be the true continuum between the C IV and He II lines, so the He II profile is well defined on its blue wing. However, there is significant overlap with O III] λ1664 on the red wing. We fit each of the O III] members with the N IV] template profile in the correct intensity ratio, and the residual from that fit was then adopted as our He II template.

Figure 3 shows all of the important line profiles from J1254+0241, lined up in velocity space. The differences between the template profiles are modest, but we believe that they are real. We then proceeded to fit each of the other lines, trying in turn each of our template profiles to see which one fit the best. In the case of unraveling the blends involving several lines (cf. Lyα + N V, and the λ1909 blend), this included trial fits with several different combinations of the templates. Whenever an observed line is a multiplet with significant wavelength spacing between the members (*i.e.* N V λ1240, Si II λ1263, C IV λ1549, O III] λ1664, N III] λ1751 and Al III λ1858), we first used the template profile to construct an artificial blend with the different members having the relative fluxes predicted by the LOC model described in §5, and then we fit the whole artificial blend to the observed profile. In such cases, the fluxes listed in Table 2 are for the entire multiplet.

Figure 4 shows the best fits for several of the blends, to illustrate the accuracy achieved. Note that the N III] λ1751 blend has another broad, flat-topped emission feature connecting onto its red wing, which we made no attempt to fit.

**3.2 J1546+5253**

We eventually identified three distinctly different template profiles in this object. Many of the emission lines (including C IV λ1549 and O III] λ1664, which are important for our abundance analysis) are chopped up by overlying narrow features from the $z = 0.792$ absorption line system found by Prochter et al. (2006), making several of our intensity measurements particularly uncertain.

The best-defined template profile is C II λ1335.1. It is narrow (FWHM = 1700 km s$^{-1}$) and reasonably strong, with $I$(C II)/$I$(C IV λ1549) = 0.054. This feature is actually a triplet, with the two much stronger lines separated by 1.13 Å, but we treated it as a single line and made no attempt to deblend it. The N IV] λ1486 feature is broader, with FWHM = 3700 km s$^{-1}$, and is fairly noisy. It was smoothed with a polynomial fit and used as a second template. It has a moderately asymmetrical wing extending about 4000 km s$^{-1}$ to the red (at zero intensity). Fitting the C IV profile was very problematical, because it is cut by Fe II λλ2586, 2600 and Mn II 2576 in the $z = 0.792$ absorption system. In the end, we just linearly interpolated over the absorption features to get a starting point for our iterative process for deblending the two C IV emission components. The resulting C IV template profile is similar to N IV] in its core, but has very extended wings on both sides (extending from -5000 to +10,000 km s$^{-1}$ at zero intensity). The wings are clearly real, and the agreement of the core profile with N IV] suggests that our crude interpolation produces something reasonably close to the correct shape. Figure 5 compares these templates and some of the observed profiles for J1546+5253.

The C IV template gives the best fit to the N V λ1240 blend. Both of the other templates are too narrow. The uncertainty in the N V measurement comes from not knowing the profile of Lyα. For our best fit, we fit a C IV template at the position of Lyα, but matched the observed flux only over the wavelength range λλ1220-1230Å. This means that for the purpose of measuring N V, we assumed that Lyα has the same broad red wing as does C IV. However, subtracting this Lyα fit leaves behind a strong central Lyα core.

The upper limit on the N V doublet's strength comes from setting the Lyα strength to zero and fitting the observed peak at N V with an artificial blend made from two C IV profiles having the correct intensity ratio. The minimum for the N V strength was found by subtracting off the artificial doublet blend until the bump caused by N V on the observed Lyα wing just barely disappeared. For all of these fits, we also included the Si II λ1263 multiplet represented by a synthetic blend made from the C IV template.

The "best" flux measurement given in Table 2 for Lyα comes from the residual spectrum after fitting N V and Si II with their best estimates.

We separated out the He II profile by fitting the N IV] template to the O III] λ1664 multiplet with which He II is blended. However, the peak of the O III] feature is cut up by the Mg II λλ2796, 2803 doublet in the $z = 0.792$ absorption system, so the fit to O III] is highly uncertain. We used the higher resolution spectrum for this, because the absorption lines are better separated from each other so a wavelength point that is possibly free of absorption can be seen between them. The minimum and maximum values listed for O III] in Table 2 are for the case where the template profile did not and did include this point, respectively. There is a factor of two difference between them. For the best O III] flux value, which we consider to be highly dubious, we averaged the minimum and maximum values. The He II strength was then measured by summing the remaining flux after subtracting off this "best" fit to O III].

The N III] λ1751 blend was fitted by constructing an artificial blend of N IV] template profiles in the same way that was done for J1254+0241. We found that the fit was significantly better (i.e. the peaks lined up) when the artificial blend was shifted to the blue by 460 km s$^{-1}$. The minimum flux quoted in Table 2 for this multiplet is for no velocity shift, and the maximum value is for a more modest velocity shift of -310 km s$^{-1}$.

In addition to having the strong N III] λ1751 feature which originally drew our attention to this object, the spectrum of J1546+5253 is unusual in several other respects. There is an exceptionally large range in line widths, from 1700 km s$^{-1}$ for the C II λ1335 line, to 5500 km s$^{-1}$ for the C IV λ1549 line. There is a strong, narrow feature at λ1787, which we identify as Fe II multiplet UV 191. It has a far greater strength relative to other Fe II lines than in our Fe II template (the template was fitted to the blended Fe II "bump" in the λλ2240–2650Å range, which is not very strong in this object). Finally, strong Fe III multiplet UV 34 lines are blended with Si III] λ1892 and C III] λ1909 (these are not included at all in our Fe II template). The combination of these peculiarities immediately brought to mind the spectra of the two unusual QSOs H0335-336 (Hartig & Baldwin 1986; Baldwin et al. 1996) and I Zw 1 (Laor et al. 1997). The first of these objects is a low-excitation Broad Absorption Line (BAL) QSO with very narrow emission lines, while I Zw 1 is a low redshift AGN which is famous for its strong, narrow Fe II emission spectrum (and which is the source of the Fe II template used here). Figure 6 compares the spectrum of J1546+5253 to that of I Zw 1. Figure 6 and also Figure 1 show the spectrum of J1546+5253 both before and after subtracting the Fe II template; the amount subtracted off is very small because the λλ2240–2650 bump is quite weak. It is clear from Figure 6 that the λ1909 blend in J1546+5253 includes Fe III λλ1914.06, 1926.03 with about the same intensity ratio relative to C III] λ1909 as in the spectrum of I Zw 1. Both the Fe II UV 191 multiplet and N III] λ1751 blend are as strong in I Zw 1 as they are in J1546+5253, so all of these unusual low-ionization lines appear to have been strengthened by about the same factor in both QSOs.

The C III] and the Fe III lines in the spectrum of J1546+5253 are quite narrow, and only the C II template profile gives a good fit. However, Si III] λ1892 appears to be broader, with the N IV] template giving the best fit. It is possible to distinguish the individual peaks of C III] λ1908.73 and the Fe III λ1914.06 lines, and the two lines have a roughly 1:1 flux ratio, just as is seen in I Zw 1. This permits a reasonably accurate measurement of the strength of the C III] line. The strength of Si III] λ1892.03 is not nearly as well measured, because due to its broader profile it is impossible to separate it from the Fe III λ1895.46 line. The "best" flux value given in Table 2 for Si III] comes from the χ-squared minimization process, which assigned a relatively large flux to Si III] at the expense of finding a weaker Fe III λ1895.46 line.

The resulting Fe III λ1895.46/λ1914.06 flux ratio is about three times weaker than was measured by Baldwin et al. (1996) for H0335-336. If we arbitrarily set the strength of λ1895.46 to be the same relative to λ1914.06 as it is in H0335-336, then the Si III] line in J1546+5253 would be two times weaker than the "best" value listed in Table 2; we list this as the minimum value for Si III]. The maximum value for Si III] comes from setting the strength of Fe III λ1895.46 to zero, which is unlikely. The true strength of Si III] could be nearly a factor of two different in either direction from the best fit value and still lie within the acceptable range.

Figure 7 shows the best fits to the same four spectral regions for J1546+5253 that were shown in Figure 4 for J1254+0241.

### 3.3 J0909+5803 and J1550+0236

Although we did not attempt to analyze the spectra of these two objects, we note here that J0909+5803 is a BAL QSO (Trump et al. 2006). N V and C IV are badly affected by their own BAL troughs, and the C IV trough also cuts across the N IV] emission line. This object is therefore not suitable for the type of abundance analysis carried out in the next section. The spectrum of J1550+0236 looks much like those of J1254+0241 and Q0353-383, with N V, N IV] and N III] all stronger than usual compared to the other emission lines. However, all of the lines in this QSO have lower equivalent widths than in J1254+0241 and Q0353-383, and this object is also about a magnitude fainter, so we were not able to get a spectrum with sufficient *S/N* ratio for a useful analysis.

### 4. Abundance Analysis

We apply the same abundance analysis procedure here that was used by Baldwin et al. (2003b) for Q0353-383. This uses the observed intensity ratios of nitrogen lines to cooling lines of other elements, and compares them to the ratios predicted by the Locally Optimally-Emitting Cloud (LOC) model of the BELR for a segmented power law, from Hamann et al. (2002).

Table 3 lists the diagnostic line ratios for J1254+0241 and J1546+5253, and also for comparison gives the values found by Baldwin et al. (2003b) for Q0353-383. Table 4 gives the metallicity *Z* indicated by each of these line ratios, relative to the solar value. As was done in Baldwin et al. (2003b), we have adjusted the Hamann et al. predictions to the revised solar C, O and Fe abundances found by Allende Prieto et al. (2001, 2002) and Holweger (2001). This modification means that the same N V (relative to other strong lines) now occurs at about 30% lower *Z* than before, which we accounted for by subtracting 0.11 from the log $Z/Z_\odot$ values given by Hamann et al. The three panels of Figure 8 then plot these observed metallicities, separately for each QSO, onto the curves of intensity ratio vs. metallicity predicted by the LOC model.

### 5. Discussion

Previous studies (Shemmer & Netzer 2002; Dietrich et al. 2003) have shown that in large samples of QSOs the N IV]/C IV ratio indicates considerably lower metallicity than does the N V/C IV ratio, while the N IV]/O III] ratio gives abundances in reasonable agreement with those found from N V/C IV, N V/O VI, N V/(O VI + C IV) and N III]/O III]. Dietrich et al. (2003) also found that the N III]/C III] and N V/He II ratios are not "robust" abundance indicators. The same pattern is seen in the intensity ratios shown on Figure 8. The N IV]/C IV, N III]/C III] and N V/He II values are indicated by, respectively, the filled and open squares and the triangles. If these points are disregarded, the remaining points give reasonably tight abundance agreement internally for each of the three QSOs.

The results for J1254+0241 are very similar to those found previously for Q0353-383. The robust line ratios all point to similarly high metal enrichment. The metallicity is about 1.0 dex above solar in J1254+0241, as compared to about 1.2–1.3 dex above solar in Q0353-383. In both of these objects, the metallicities derived from N III] and N IV] strengths agree well with those derived from N V strengths. This again validates the use of N V for abundance measurements in QSOs.

J1546+5253 appears to have lower metallicity, with log($Z/Z_\odot$) ~ 0.6–0.7. It's spectrum also shows a greater scatter between the robust abundance indicators than in the other two objects. We note that of the lines used in the diagnostic intensity ratios for this object, N V, N IV], He II, N III] and C III] are reasonably well measured, but the C IV and O III] fluxes are much more uncertain. Therefore, even the robust abundance indicators N V/C IV, N IV]/O III] and N III]/O III] are not measured as well in this object as they are in J1254+0241 or Q0353-383. However, we do not believe that this entirely explains the very large scatter in the abundances derived for J1546+5253 from different line ratios.

Different line ratios will indicate different abundances if the parameters in the LOC model used by Hamann et al. (2002) do not correctly describe the QSO in question. The idea behind the LOC model (Baldwin et al. 1995) is that since we know (from reverberation measurements and line profile differences) that the broad emission lines do not all come from one spot, it is better to use an approximate description of an extended BELR than no description at all. The LOC models assume a power-law distribution of the ionizing photon flux $\Phi_H$ striking each portion of the BELR, and the gas density $n_H$. The standard power law indices used by Hamann et al. do a good job of reproducing the emission line intensity ratios in the spectra of typical QSOs. The fact that these standard parameters also give self-consistent results for the metal enrichment measurements in J1254+0241 and Q0353-383 implies that the gas distribution in the $\Phi_H - n_H$ plane in these two objects is similar to that in typical QSOs with lower metallicities.

J1546+5253 shows a much wider scatter in the metallicities derived from the different indicators. Since the intensity ratios for this object are unusual, we would deduce that the distribution of the BELR gas on the $\Phi_H - n_H$ plane is also unusual. The Si III]/C III] ratio in this object is higher than usual, which could be interpreted as indicating that the average BELR density in this object is abnormally high, as is also found in I Zw 1 (Laor et al. 1997). In the LOC model this would again be stated as a difference in the way the $\Phi_H - n_H$ plane is covered, with a higher fraction of the BELR gas having $n_H \geq 10^{11.5}$ cm$^{-3}$ where strong Si III] but very little C III] emission is produced (see Fig 3e of Korista et al. 1997).

The emission lines from J1546+5253 also are notable for showing very large differences in the widths and asymmetries of different lines even of similar ionization level. The most discrepant abundance is for the N IV]/C IV ratio, and there is also a considerable mismatch between the N IV] and C IV profiles, since the N IV] profile does not include the broad wings on the C IV profile. The C III] and Fe III profiles come from a region with very narrow lines, while Si III] apparently comes from a somewhat broader-lined part of the BELR. Although we fitted the N III] profile with the N IV] template, we had to shift the template by a fairly large velocity difference (-460 km s$^{-1}$), showing that the profile agreement is fortuitous and the lines from these two adjacent ionization states of N come from quite different parts of the BELR. The LOC model is supposed to remove such problems by accounting for emission from throughout the BELR, but only to the extent that the assumed $\Phi_H$ and $n_H$ distributions correctly describe the clouds from which we actually can see emission lines.

All of the unusual features in the spectrum of J1546+5253 could be just a viewing angle effect. The spectra of J1546+5253 and also I Zw 1 have several similarities to those of LoBAL QSOs (as represented by H0335-336). The viewing aspect angle is thought to be at least one key factor in defining BAL QSOs. Perhaps all of these objects, including J1546+5253, are cases where we are seeing winds coming off a nearly edge-on accretion disk. In addition to accentuating the velocity differences between different gas components, this orientation could lead to some of the BELR gas being shielded from our view, producing the unusual line intensity ratios. Searches through the SDSS data base for QSOs with very strong lines of N III] or Fe II UV 191 will be a good way to efficiently discover this type of objects and better understand the details of the structure of the BELR in the general population of QSOs. However, we also note that in spite of its strong Fe II UV 191 emission, J1546+5253 does not have strong Fe II emission in general, which is quite different from I Zw 1 and H0335-336, which both have very strong emission in the Fe II UV bump.

Returning to the question of robust vs. non-robust abundance indicators, the N V/He II ratio is thought to be non-robust because it is sensitive to the shape of the ionizing continuum (see Fig. 5 of Hamann et al. 2002). For Q0353-383 and J1254+0241, the metallicity derived from N V/He II agrees well with that derived from N V/C IV, so we deduce that the segmented power law used to calibrate the x-axis of Figure 8 is a reasonable description of the ionizing continuum shape in those two objects. For J1546+5253, the N V/He II ratio is about 0.26 dex larger than is predicted for the metallicity indicated by N V/C IV and the other robust lines. This is about what would be expected if the continuum shape in this object is closer to that of the $\alpha = -1.0$ power law used as an alternative continuum by Hamann et al. However, no firm conclusion can be made for J1546+5253 because of the problems with the C IV measurement caused by the intervening absorber.

It is sometimes suggested that the N V $\lambda1240$ line is significantly enhanced through pumping by photons scattered by the broad wing of the Ly$\alpha$ line in outflowing BAL gas (Wang, Wang & Wang 2006; Krolik & Voit 1998). This would mean that that N V should not be used in abundance analyses such as the one here, because our photoionization models do not properly account for its formation. Hamman, Korista & Morris (1993) have given both observational and theoretical arguments showing that this N V enhancement is not a major effect. The observations of J1254+0241 and Q0353-383 also argue against such an enhancement, since the line ratios involving N V in the numerator do not give greatly higher metallicities than the ratios involving intercombination lines from other ionization states of nitrogen. It appears that in these two objects (and therefore presumably also in other objects for which the standard LOC model is a reasonable fit) the N V lines *are* properly accounted for in the photoionized models, and the N V lines *can* be used to determine metal abundances. In the case of J1546+5253, which has many spectroscopic similarities to BAL objects such as H0335-336, different line ratios involving N V indicate both the highest and lowest metallicities, so again there is no particular evidence supporting the pumping process.

Q0353-383 and (at a more modest level) J1254+0241 stand out as having extremely high metallicities in comparison to the QSO population in general, which itself has super-solar metallicity on average. In comparing the metallicities derived here to those published for much larger samples by Dietrich et al. (2003), Warner et al. (2003, 2004) and Nagao et al. (2006), it is important to note that those other works have all retained the old solar abundance scale used by Hamann et al. (2002), which gives metallicty $Z$ values approximately 30 percent higher than we find here. Allowing for that factor, or by directly comparing line ratios where given in those papers and also in the work of Shemmer & Netzer (2002) and Shemmer et al. (2004), we find that by almost all measures the strengths of N V, N IV] and N III] relative to lines of other elements for Q0353-383 lie about 0.2–0.5 dex above the highest values in the larger surveys. The exceptions are that the N V/He II and N V/O VI ratios are about the same for Q0353-383 as for the very highest individual QSOs in the Dietrich et al. (2003) sample or in various composites corresponding to the luminosities or black hole masses or accretion rates for which the metallicity is highest. Q0353-383 still seems to be an extreme object in terms of its metallicity. J1254+0241 has a lower metallicity than Q0353-383, but one that still is comparable to the highest metallicities found in the larger samples.

The next step is to use these super metal rich QSOs to try to understand what is happening in the host galaxies. The BELR in a typical QSO consists of $10^{3-4}$ M$_\odot$ or more of enriched gas. In a realistic chemical enrichment model this would be associated with a stellar population of at least $10^{4-5}$ M$_\odot$, and more likely with a bulge-sized mass of stars (Baldwin et al. 2003a). Chemical enrichment models appropriate for the central cores of massive galaxies (e.g. Hamann & Ferland 1993; Friaca & Terlevich 1998; Romano et al. 2002) terminate at $Z \sim 10$–$20$ Z$_\odot$ because of gas depletion. This will also shut off the QSO activity, after some characteristic timescale for the gas to accrete onto the black hole. The phase of rapid enrichment and QSO activity can be prolonged and reach higher metallicities if there is a source of new gas feeding into the central region, due either to infall from outside the host galaxy, or to gas feeding in from the outer regions of the same galaxy (for example, because of a bar instability). Presumably some QSOs are caught part of the way through the rapid enrichment phase in their particular host galaxy, while others are seen at the very end of the process. Chemical enrichment models provide information about the timescales involved, which will allow us to correct for this effect. The statistics of metallicities in QSO BELRs therefore will tell us about the history of star formation, chemical enrichment and the feeding of additional gas into central cores in the host galaxies. The QSOs with the very highest metallicity will tie down the endpoint of rapid chemical enrichment, while the existing (and future) studies of large numbers of more typical QSOs will provide the statistical context for how these extreme objects fit within the broader population of QSOs and host galaxies.

## 6. Conclusions

We have made very detailed studies of the spectra of two of the QSOs found by Bentz et al. (2004) to have the strongest NIII] and N IV] lines of all the QSOs in the SDSS DR1 catalogue. These represent the very highest-metallicity end of the distribution of QSO metallicities that has been studied in larger sets of individual objects and by means of composite spectra.

The spectrum of the QSO J1254+0241 shows a pattern of unusually strong nitrogen emission lines, from $N^{+4}$, $N^{+3}$ and $N^{+2}$, that is entirely consistent with this object having a N/O abundance ratio about 10 times the solar value. This is very similar to the situation for Q0353-383, another QSO with unusually strong nitrogen lines which also can be explained by an N/O ratio much higher than solar (N/O ~ 15× solar in that case; Baldwin et al. 2003b). In the picture in which high N/O ratios are due to secondary nitrogen enrichment in the CNO process, these high N/O ratios imply an overall metallicity $Z$ which is increased by the same factor above the solar values (Hamann & Ferland 1993, 1999). There are now two well-studied cases which demonstrate that the nitrogen emission lines are good abundance indicators even in the QSOs which are among the most metal-rich known.

The other QSO studied here, J1546+5253, has a lower metallicity, with $Z \sim 5Z_\odot$. The various abundance indicators in its spectrum show a much wider scatter in $Z$ than do those for J1254+0241 or Q0353-383. The spectrum of J1546+5253 is unusual in other respects as well. We suggest that this is because the particular parameters used in the "standard" LOC model employed in the analysis do not provide a good description of the distribution of the BELR gas in this particular QSO. The immediate clue that this is the case is that, in comparison to most QSOs, the integrated BELR spectrum from this object shows unusual intensity ratios for several low-ionization emission lines. This suggests that when line ratios such as N V/C IV are used to determine metallicities for large samples of QSOs, the objects with unusual low-ionization spectra similar to that of J1546+5253 should first be discarded.

### Acknowledgements

We thank Gary Ferland, Fred Hamann, Kirk Korista and the anonymous referee for some very helpful comments and suggestions. ND and JAB gratefully acknowledge support from NSF through grant AST-0305833 and from NASA through HST grant GO09885.01-A. MCB is supported by a Graduate Fellowship of the National Science Foundation.

| Table 1 |||||
|---|---|---|---|---|
| Objects Observed and Exposure Times |||||
| Object (SDSS J) | $m_i$ | z | Exposure times (s) ||
| 090902.55+580356.9 | 18.986 | 1.793 | Low res: 1×1200, 6×1800, 1×900 (wide slit) ||
| 125414.27+024117.5 | 18.319 | 1.839 | Low res: 4×1800, 3×900 (wide slit) ||
|  |  |  | High res: 6×1800, 1×900 (wide slit) ||
| 154651.75+525313.1 | 19.057 | 2.006 | Low res: 1×1200, 5×1800, 1×120 (wide slit) ||
|  |  |  | High res: 1×900, 6×1800, 1×900 (wide slit) ||
| 155007.07+023607.6 | 20.064 | 2.366 | Low res: 1×1200, 6×1800, 1×1200 (wide slit) ||

| Table 2 |||||||||
|---|---|---|---|---|---|---|---|---|
| Measured Emission Line Strengths Relative to C IV λ1549 |||||||||
|  | J1254+0241 |||| J1546+5253 ||||
| Line[1] | Best | Min | Max | Template | Best | Min | Max | Template |
| Lyα λ1215.67 | 4.11 | 3.26 | 4.52 | C IV | 3.19 | 3.04 | 3.72 | sum |
| Lyα blend[2] | 5.90 |  |  |  | 3.90 |  |  |  |
| N V λ1240 | 1.50 | 1.44 | 1.57 | C IV | 0.38 | 0.31 | 0.61 | C IV |
| Si II λ1263 | 0.16 | 0.10 | 0.30 | C IV | 0.15 | 0.07 | 0.20 | C IV |
| N IV] λ1486.50 | 0.23 | 0.23 | 0.27 | N IV] | 0.05 | 0.05 | 0.06 | N IV] |
| N IV] total[2] | 0.24 |  |  |  | 0.05 |  |  |  |
| C IV λ1549 | 1.00 | 1.00 | 1.09 | C IV | 1.00 | 0.86 | 1.00 | C IV |
| C IV total[2] | 1.02 |  |  |  | 0.99 |  |  |  |
| He II λ1640.42 | 0.23 | 0.16 | 0.23 | He II | 0.06 | 0.06 | 0.06 | He II |
| O III] λ1664 | 0.16 | 0.13 | 0.16 | N IV] | 0.08[3] | 0.06 | 0.12 | N IV] |
| N III] λ1751 | 0.40 | 0.37 | 0.42 | He II | 0.15 | 0.15 | 0.16 | N IV][4] |
| N III] total[2] | 0.43 |  |  |  | 0.16 |  |  |  |
| Al III λ1858 | 0.14 | 0.11 | 0.14 | C IV | --- | --- | --- | --- |
| Si III] λ1892.03 | 0.15 | 0.14 | 0.16 | N IV] | 0.16 | 0.08 | 0.21 | N IV] |
| Fe III λ1895.46 | --- | --- | --- | --- | 0.03 | 0.00 | 0.08 | C II |
| C III] λ1908.73 | 0.42 | 0.42 | 0.46 | C IV | 0.12 | 0.12 | 0.15 | C II |
| Fe III λ1914.06 | --- | --- | --- | --- | 0.10 | 0.09 | 0.10 | C II |
| Fe III λ1926.30 | --- | --- | --- | --- | 0.06 | 0.06 | 0.06 | C II |
| λ1909 blend[2] | 0.78 |  |  |  | 0.59 |  |  |  |
| $10^{-15}$ F(C IV) (erg cm$^{-2}$ s$^{-1}$) | 2.65 |  |  |  | 2.17 |  |  |  |

Notes
[1]Wavelengths are given to two decimal places where individual lines have been measured. Wavelengths with no decimal places indicate that the flux is measured for an entire multiplet.
[2]Sum of total flux above continuum.
[3]Profile badly affected by narrow absorption lines. See text for explanation of quoted value.
[4]For best fit, N IV] was shifted by -460 km s$^{-1}$. Quoted minimum value is for no velocity shift.

| Table 3 Log (Intensity Ratio) for Different Metallicity Indicators | | | | | | | | | |
|---|---|---|---|---|---|---|---|---|---|
| **Line Ratio** | Q0353-383 | | | J1254+0241 | | | J1546+5253 | | |
| | **Best** | **Min** | **Max** | **Best** | **Min** | **Max** | **Best** | **Min** | **Max** |
| N III]/C III] | 0.09 | --- | --- | -0.02 | -0.10 | 0.00 | 0.12 | -0.02 | 0.13 |
| N III]/O III] | 0.67 | --- | --- | 0.40 | 0.36 | 0.51 | 0.26 | 0.08 | 0.40 |
| N IV]/O III] | 0.48 | --- | --- | 0.15 | 0.15 | 0.32 | -0.20 | -0.37 | -0.06 |
| N IV]/C IV | -0.52 | --- | --- | -0.64 | -0.68 | -0.56 | -1.28 | -1.28 | -1.19 |
| N V/He II | 0.98[1] | 0.86 | 1.10 | 0.82 | 0.80 | 1.00 | 0.78 | 0.69 | 0.99 |
| N V/C IV | 0.29[1] | 0.17 | 0.41 | 0.18 | 0.12 | 0.20 | -0.42 | -0.50 | -0.15 |
| N V/O VI | 0.33[1] | 0.21 | 0.45 | --- | --- | --- | --- | --- | --- |
| N V/(C IV+O VI) | 0.01[1] | -0.11 | 0.13 | --- | --- | --- | --- | --- | --- |

[1] Logarithmic average of minimum and maximum values.

| Table 4 Metallicity Values for Different Metallicity Indicators | | | | | | | | | |
|---|---|---|---|---|---|---|---|---|---|
| **Log $Z/Z_\odot$** | Q0353-383 | | | J1254+0241 | | | J1546+5253 | | |
| | **Best** | **Min** | **Max** | **Best** | **Min** | **Max** | **Best** | **Min** | **Max** |
| N III]/C III] | 0.94 | --- | --- | 0.75 | 0.58 | 0.79 | 0.97 | 0.74 | 0.99 |
| N III]/O III] | 1.28 | --- | --- | 0.87 | 0.83 | 1.03 | 0.72 | 0.52 | 0.86 |
| N IV]/O III] | 1.42 | --- | --- | 1.00 | 1.00 | 1.21 | 0.66 | 0.43 | 0.78 |
| N IV]/C IV | 0.79 | --- | --- | 0.70 | 0.67 | 0.76 | 0.09 | 0.09 | 0.18 |
| N V/He II | 1.26[1] | 1.12 | 1.39 | 1.08 | 1.05 | 1.28 | 1.04 | 0.94 | 1.27 |
| N V/C IV | 1.36[1] | 1.18 | 1.54 | 1.19 | 1.10 | 1.21 | 0.44 | 0.33 | 0.75 |
| N V/O VI | 1.25[1] | 1.08 | 1.42 | --- | --- | --- | --- | --- | --- |
| N V/(C IV+O VI) | 1.30[1] | 1.14 | 1.46 | --- | --- | --- | --- | --- | --- |

[1] Best values are calculated from the logarithmic average of minimum and maximum values of the line ratios.

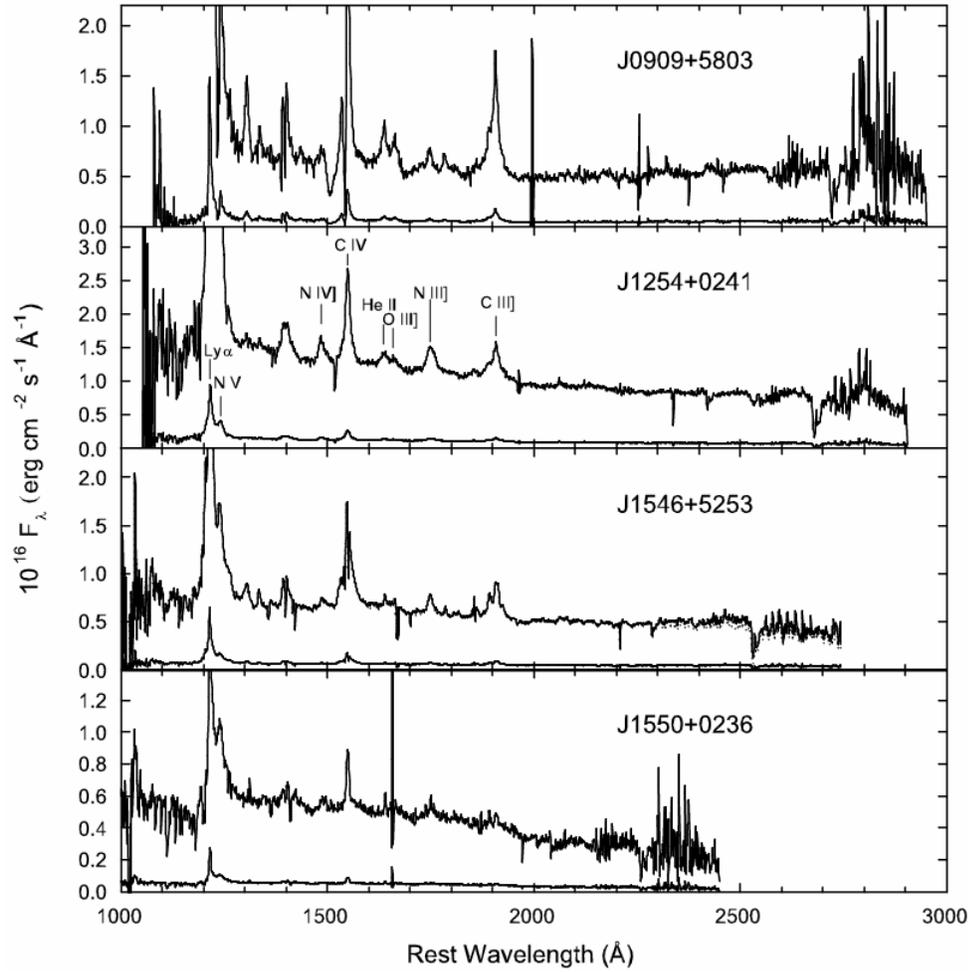

Figure 1. Low-resolution spectra, after reddening correction and with the wavelengths converted to the rest frame. The lower plot in each panel shows the same spectrum with its flux divided by 10. Ly$\alpha$ and the emission lines used in the abundance analysis are marked. The dotted line barely visible beneath the J1546+5253 spectrum shows the result of subtracting the Fe II template.

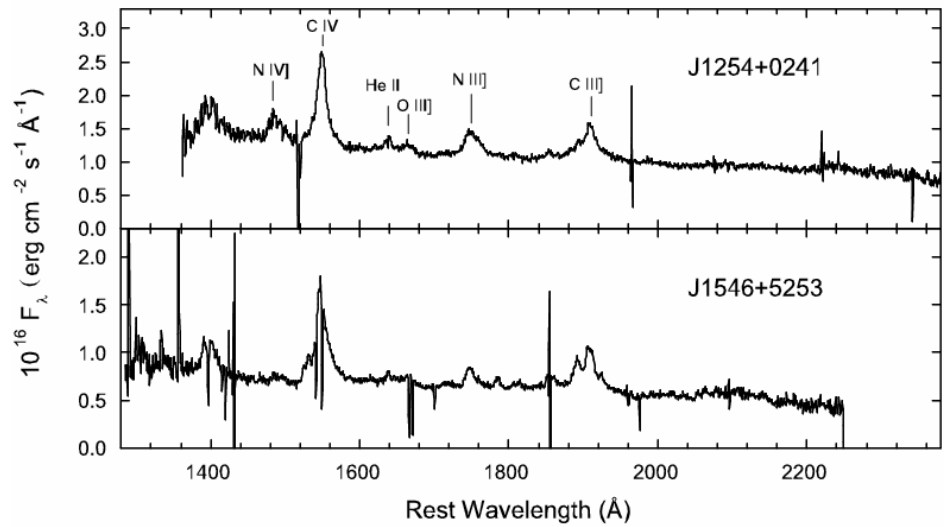

Figure 2. Higher-resolution spectra, after reddening correction and with the wavelengths shown in the rest frame.

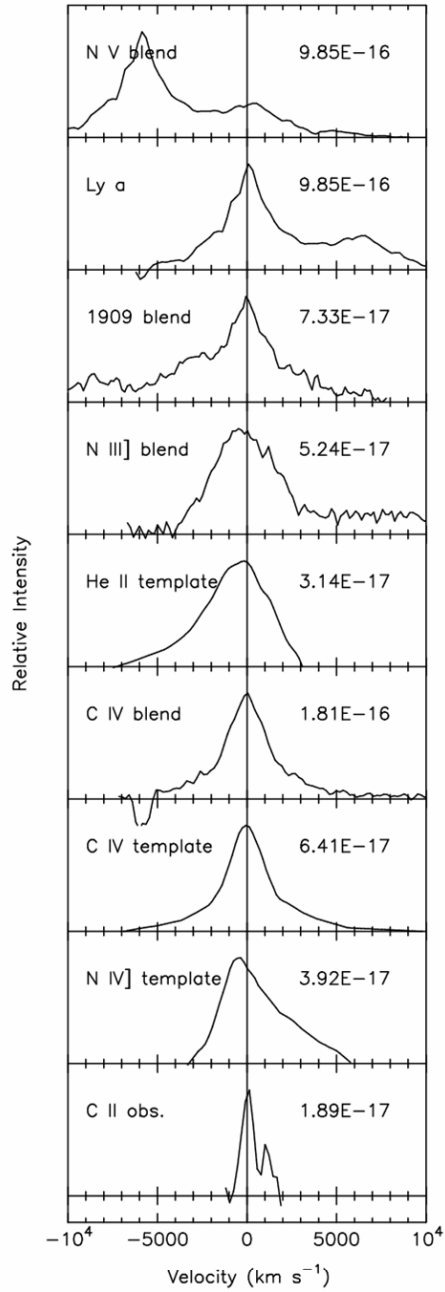

Figure 3. Velocity profiles of the stronger emission lines in J1254+0241. The numbers in the upper right corner of each panel show the value of the peak line flux in each panel, as an indication of the relative line intensities.

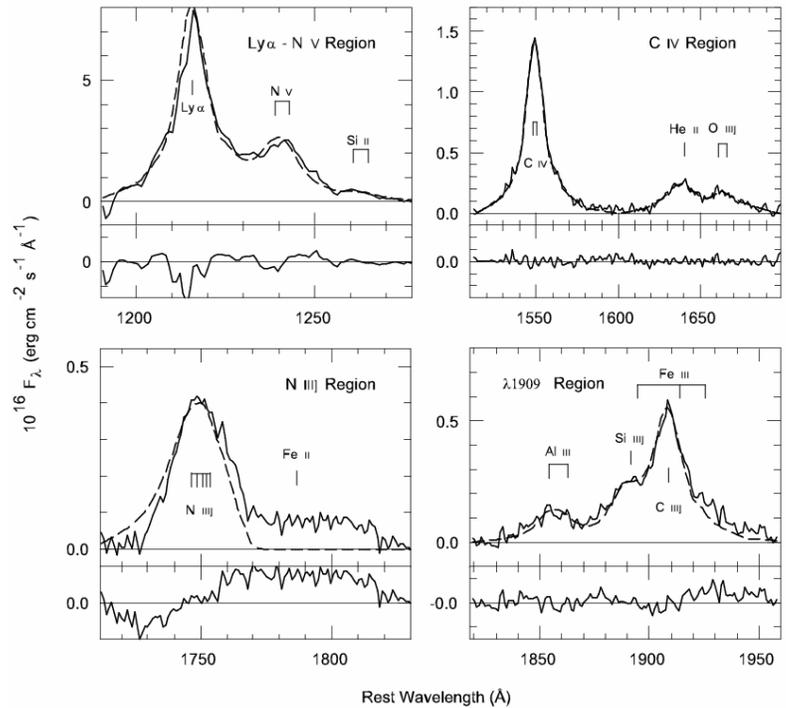

Figure 4. Line profile fits to four blended regions of the spectrum of J1254+0241. In the upper panel in each box, the solid line shows the observed spectrum and the dashed line is the best fit. The lower panel in the box shows the residual, on the same vertical scale. The tick marks show the positions of the individual lines included in the synthetic line profiles.

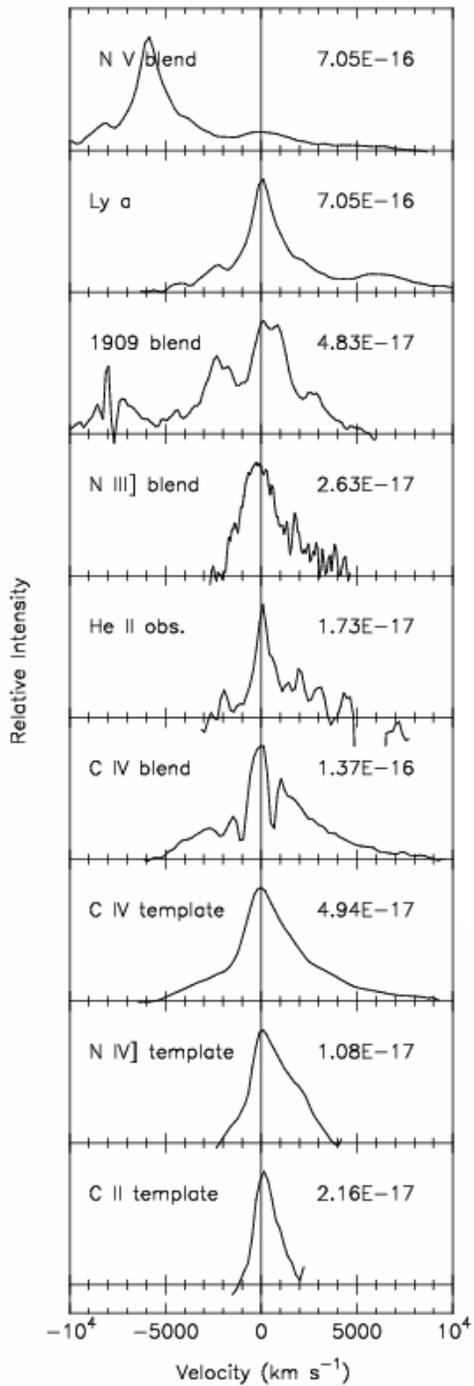

Figure 5. Velocity profiles of the stronger emission lines in J1546+5253, with notation as in Figure 3.

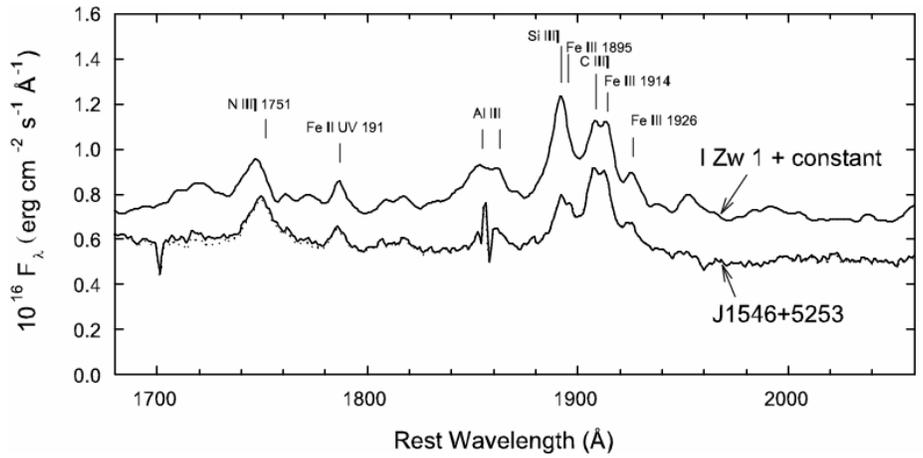

Figure 6. Comparison of the spectra of J1546+5253 (lower line) and I Zw 1 (upper line). The two spectra first were scaled so that their continuum heights at λ2100Å matched, then the spectrum of I Zw 1 was shifted upwards by adding a constant. The I Zw 1 data are from Laor et al. (1997), and have been modestly smoothed (using a 540 km s$^{-1}$ FWHM Gaussian) so that they would have the same width as the lines in J1546+5253. The dotted line shows the J1546+5253 spectrum after subtracting the Fe II template.

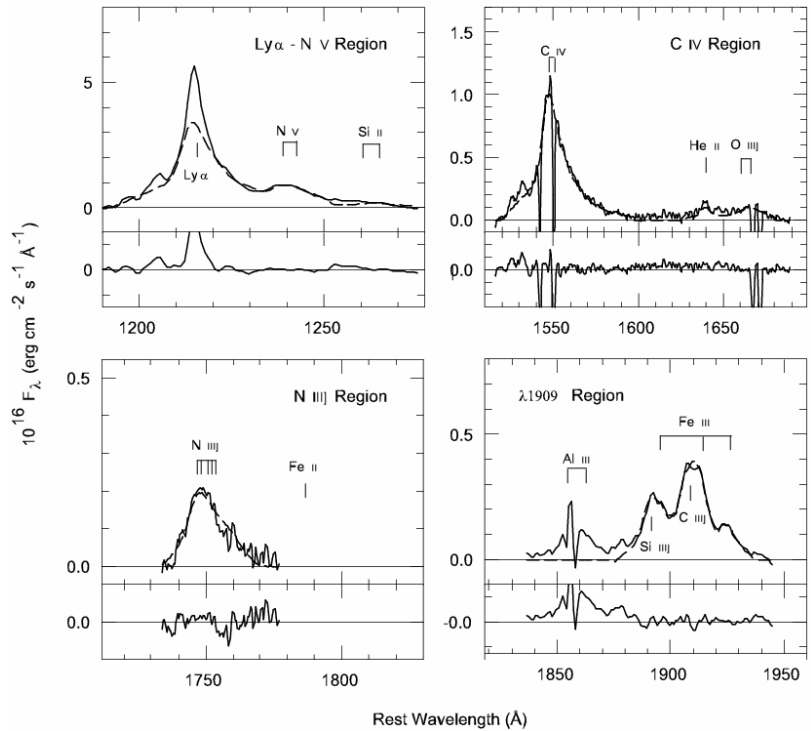

Figure 7. Line profile fits to four blended regions of the spectrum of J1546+5253, plotted in the same way as in Fig. 4.

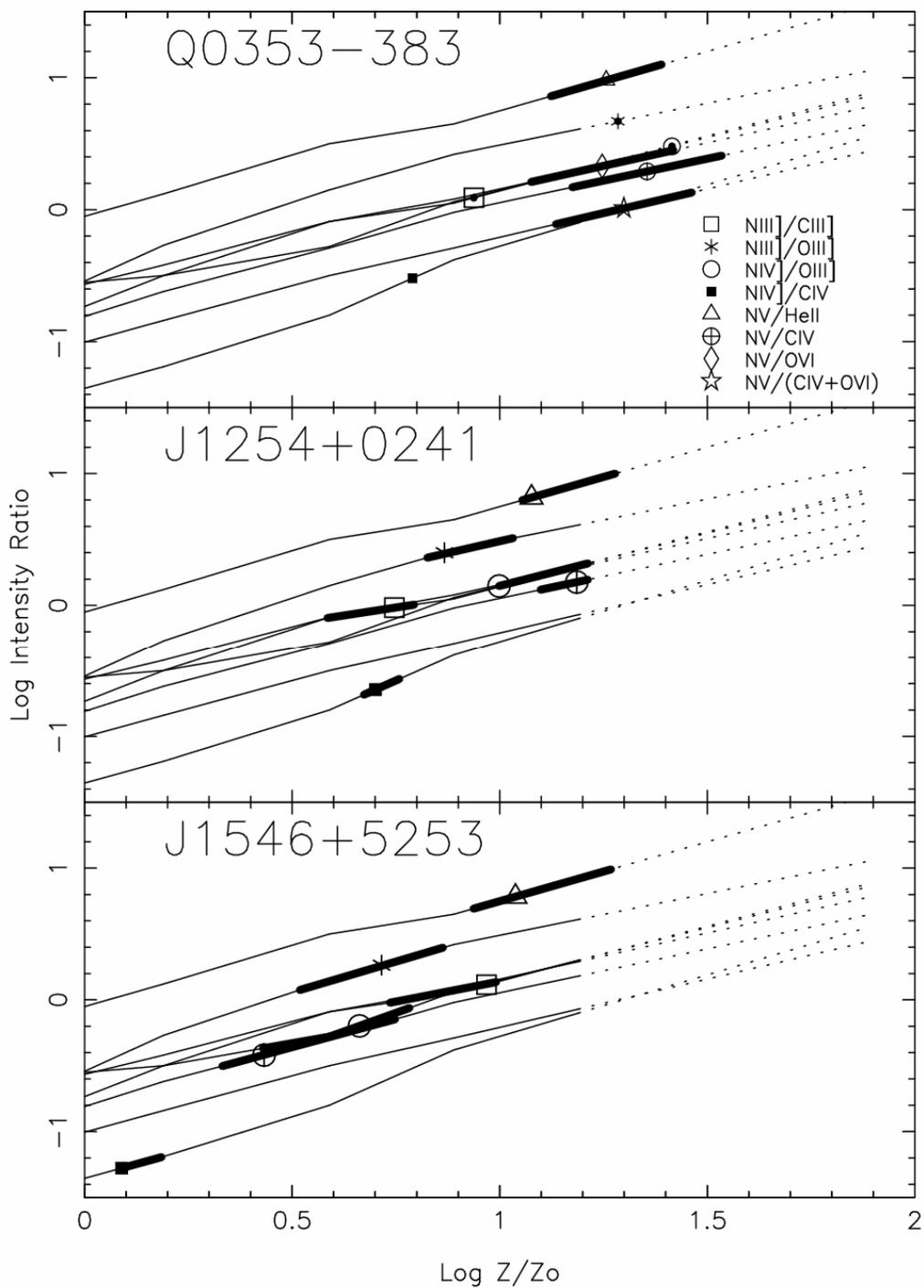

Figure 8. Abundance analysis results. The thin solid curves are the intensity ratios as a function of metallicity *Z* as predicted by the LOC models. The dotted portions of these curves are extrapolations beyond the highest *Z* for which a model was run. The symbols show the best measurements of each line ratio, and the heavy lines show the range of uncertainty corresponding to the minimum and maximum values listed in Table 4.

18